\documentstyle [prb,aps]{revtex}

\begin{document}
\draft
\title{{\bf Full Symmetry of
Single- and Multi-wall Nanotubes}}
\author{M. Damnjanovi\'c\cite{EmailY}, I. Milo\v sevi\'c, T. Vukovi\'c
and  R. Sredanovi\'c}
\address{Faculty of Physics, University of Beograd,
http//:www.ff.bg.ac.yu\\ POB 368, Beograd 11001, Yugoslavia}
\maketitle
\date{\today}
\begin{abstract}
The full symmetry groups for all single- and multi-wall carbon nanotubes
are found. As for the single-wall tubes, the symmetries form nonabelian
nonsymorphic line groups, enlarging the groups reported in literature.
In the multi-wall case, any type of the line and the axial point groups
can be obtained, depending on single-wall constituents and their
relative position.
Several other consequences are discussed: quantum numbers and related
selection rules, electronic and phonon bands, and their degeneracy,
application to tensor properties.
\end{abstract}
\pacs{61.46.+w,02.20}
\section{Introduction}
The single-wall carbon nanotubes are quasi 1D cylindrical structures
\cite{IIJIMA,TUTORIAL,SFCN}, which can be imagined as rolled up
cylinders of the 2D honeycomb lattice of the single atomic layer of
crystalline graphite. Frequently, several single-wall tubes are
coaxially arranged, making multi-wall nanotube. Since their diameters
are small (down to $0.7\,\rm{nm}$) in comparison to lengths (up to tens
of $\mu m$), the theoretical model of the extended (i. e. infinite, and
hence without caps at the ends) nanotube is well justified.

The symmetry of the nanotubes is relevant both for deep insight into the
physical properties (quantum numbers, selection rules, optical activity,
conducting properties, etc.) and to simplify calculations. As for the
single-wall tubes, symmetry studies started by the classification of the
graphene tubes according to fivefold, threefold or twofold axis of the
related $C_{60}$ molecule \cite{C-60}, and gave just a part of their
point group symmetry. The translational periodicity was discussed in
context of the nanotube metallic properties \cite{HAMADA}.  Finally, the
helical and rotational symmetries were found \cite{WHITE,JISHI}: the
screw axis was characterized in terms of tube parameters, as well as the
order of the principle rotational axis \cite{JISHI}. The first goal of
this paper is to complete this task, giving the full geometric symmetry
of the extended single-wall nanotubes. Due to their 1D translational
periodicity, the resulting groups are the line groups \cite{IRY,YV}; it
appears that only two line group families are relevant: the 5th for the
chiral, and the 13th for the armchair and the zig-zag nanotubes.

The symmetry of the double- and multi-wall tubes has never been
seriously studied, despite their importance for applications in
nanodevices. Here, we present the exhaustive list of symmetries of such
tubes. Depending on the single-wall constituents, and their relative
arrangement, the resulting nanotube symmetry may be either line group or
axial point group.

In section \ref{Ssn}, at first the necessary notions on the line
groups are briefly summarized, and the relevant notation is introduced.
Then, in subsection \ref{Ssswn}, the line groups of all the nanotubes
are derived: the familiar symmetries of the original graphene lattice
are transferred into the tubular geometry and those which remain
symmetries of the rolled up lattice form the corresponding line group.
Besides the rotational, translational and helical symmetries, the
horizontal axes and (for zig-zag and armchair tubes) mirror- and
glide-planes are also present.  The symmetry groups of multi-wall
nanotubes are studied in subsection \ref{Ssmwn}. Note that among them
there are also tubes being not translationally periodic.

Some of the possible applications of symmetry in the physics of
nanotubes are discussed in the last section.

\section{Symmetry of nanotubes}\label{Ssn}
The line groups \cite{IRY,IAY} contain all the symmetries of the systems
periodical in one direction and usually are used in context of
stereoregular polymers and quasi-1D subsystems of 3D crystals.  It
immediately follows that, being periodic along its axis, any extended
single-wall nanotube has the symmetry described by one of the line
groups.

All the line group transformations leave the tube axis ($z$-axis, by
convention) invariant. Consequently, such a transformation $(P|t)$
(Koster-Seitz symbol) is some point group operation $P$ preserving the
$z$-axis, followed by the translation for $t$ along the $z$-axis. Action
on the point ${\bf r}=(x,y,z)$ gives $(P|t){\bf r}=(x',y',z')$ with
\begin{equation}\label{Edefelem}
x'=P_{xx}x+P_{xy}y,\quad y'=P_{yx}x+P_{yy}y,\quad z'=P_{zz}z+t.
\end{equation}
Here, $P_{ij}$ are elements of the $3\times 3$ matrix of $P$ in the
Cartesian coordinates; those coupling $z$ to the other axes vanish.
Such point operations are called axial, and they form seven types of the
axial point groups \cite{ELDOB}: ${\bf C}_n$, ${\bf S}_{2n}$, ${\bf C}_{nh}$,
${\bf C}_{nv}$, ${\bf D}_n$, ${\bf D}_{nd}$, ${\bf D}_{nh}$, where $n=1,2,\dots$ is the
order of the principle rotational axis.

There are infinitely many line groups, since there is no
crystallographic restriction on the order of the principle axis, and
they are classified within 13 families. Each line group is a product
${\bf L}={\bf Z}{\bf P}$ of one axial point group ${\bf P}$ and one infinite cyclic
group ${\bf Z}$ of generalized translations (screw-axis ${\bf T}^r_q$, pure
translations ${\bf T}={\bf T}^0_1$, or glide plane ${\bf T}_c$, generated by the
transformations $(I|a)$, $(C^r_q|\frac{n}{q}a)$ and
$(\sigma_v|\frac{a}{2})$, respectively \cite{SCREWY}). Thus, to
determine the full symmetry of a nanotube, both of these factors (having
only the identical transformation in common) should be found. The point
factor ${\bf P}$ should be distinguished from the isogonal point group
${\bf P}_I$ of the line group \cite{IAY}: only for the symorphic groups when
${\bf Z}={\bf T}$, these groups are equal; otherwise ${\bf P}_I$ is not a subgroup
of ${\bf L}$.  Due to the convention \cite{SCREWY}, $2\pi/q$ is the minimal
angle of rotation performed by the elements of the line group (if the screw
axis is nontrivial it is followed by some fractional translation), as
well as by its isogonal point group.

The easiest way to determine the line group ${\bf L}$ of a system is to find
at first the subgroup ${\bf L}^{(1)}$, containing all the translations and
the rotations around the principle axis (including the ones followed by
fractional translations). Having the same screw axis (${\bf T}$ is a special
case) as ${\bf L}$, and the same order $n$ of the principle axis, this
subgroup ${\bf L}^{(1)}={\bf T}^r_q{\bf C}_n$ is the maximal subgroup from the first
line group family. Then the symmetries complementing ${\bf L}^{(1)}$ to
${\bf L}$ should be looked for. To complete ${\bf Z}$, it should be checked if
there is a vertical glide plane. Also, ${\bf C}_n$ is to be complemented to
${\bf P}$ by eventual additional point group generators; at most two of them
are to be chosen among the mirror planes, horizontal rotational axes of
order two or rotoreflection axis (refining pure rotations that are
already encountered in ${\bf C}_n$).

\subsection{Single-wall nanotubes}\label{Ssswn}
Elementary cell of the hexagonal honeycomb lattice (Fig. \ref{Fmed}) is
formed by vectors $\vec{a}_1$ and $\vec{a}_2$ of the length
$a_0=2.461$\AA; within its area $S_g=\sqrt{3}/2a^2_0$ there are two
carbon atoms at positions $(\vec{a}_1+\vec{a}_2)/3$ and
$2(\vec{a}_1+\vec{a}_2)/3$.  The single-wall nanotube $(n_1,n_2)$ is
formed when the honeycomb lattice is rolled up, in such a way that the
chiral vector $\vec{c}=n_1\vec{a}_1+n_2\vec{a}_2$ becomes the
circumference of the tube (its end and origin match). The tubes
$(n_1,0)$ and $(n_1,n_1)$ are called zig-zag and armchair, respectively,
while the others are known as the chiral ones. The chiral angle $\theta$
of the nanotube is the angle between the chiral vector $\vec{c}$ and the
zig-zag direction $\vec{a}_1$.  When $0\leq\theta<\pi/3$, all the tubes
are encountered; in fact, for the zig-zag and the armchair nanotubes
$\theta$ equals $0$ and $\pi/6$, respectively, and between these
chiralities lay chiral vectors of all the chiral nanotubes with
$n_1>n_2>0$ (the tubes $(n_2,n_1)$, with $\pi/6<\theta<\pi/3$, are their
optical isomers).

{\footnotesize\begin{figure}[hbt]\begin{center}
\unitlength=1.00mm
\linethickness{0.4pt}
\begin{picture}(45.00,20.00)
\put(0.00,0.00){\framebox(45.00,20.00)[cc]{FIG 1}}
 \end{picture}
\caption[]{\label{Fmed} {\bf Symmetries of the honeycomb lattice.} {\sl
For the chiral $(8,6)$, zig-zag $(6,0)$ and armchair $(6,6)$ tubes the
chiral vectors $\vec{c}$ are depicted by the arrows. The $U$ and $U'$
axes pass through the circles (perpendicular to the honeycomb). In the
zig-zag and armchair case, the bold lines $\sigma_v$ and $\sigma'_v$
represent the vertical mirror and glide planes of the lattice, being
orthogonal to $\vec{c}$; the planes parallel to $\vec{c}$ are denoted as
$\sigma_h$ and $\sigma'_h$; $U$ is the intersection of the mirror
planes, and $U'$ of the glide planes.}}
\end{center}\end{figure}
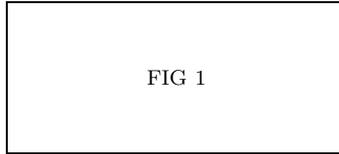}

There are $n=\text{GCD}(n_1,n_2)$ (the greatest common divisor) honeycomb
lattice points laying on the chiral vector. The translations for
$s\vec{c}/n$ in the chiral direction, on the tube appear as the
rotations for $2s\pi/n$ ($s=0,1,\dots$) around the tube axis.  Thus, the
principal axis of order $n$ is a subgroup of the full symmetry of the
tube $(n_1,n_2)$:
\begin{equation}\label{Erot}
{\bf C}_n,\quad n=\text{GCD}(n_1,n_2).
\end{equation}
Obviously, $n=n_1$ for the zig-zag $(n_1,0)$ and the armchair
$(n_1,n_1)$ nanotubes.

To the primitive translation of the tube corresponds the  vector
$\vec{a}=a_1\vec{a}_1+a_2\vec{a}_2$ in the honeycomb lattice, being the
minimal one among the lattice vectors orthogonal onto $\vec{c}$.
Therefore, $a_1$ and $a_2$ are coprimes, yielding
\begin{equation}\label{Etrans}
\vec{a}=-\frac{2n_2+n_1}{n{\cal R}}\vec{a}_1+\frac{2n_1+n_2}{n{\cal R}}\vec{a}_2,
\quad a=|\vec{a}|=\frac{\sqrt{3(n^2_1+n^2_2+n_1n_2)}}{n{\cal R}}a_0,
\end{equation}
with ${\cal R}=3$ if $\frac{n_1-n_2}{3n}$ is integer and ${\cal R}=1$ otherwise.
For the zig-zag and the armchair tubes $a=\sqrt{3}a_0$ and $a=a_0$,
respectively.  The elementary cell of the tube is the cylinder of the
height $a$ and area $S_t=a|\vec{c}|$; it contains
$\frac{S_t}{S_g}=2\frac{n_1^2+n_2^2+n_1n_2}{n{\cal R}}$ elementary graphene
cells \cite{JISHI}. So, the translational group ${\bf T}$ of the nanotube is
composed of the elements $(I|ta)$, $t=0,\pm1,\dots$

The encountered symmetries ${\bf T}$ and ${\bf C}_n$ originate from the
honeycomb lattice translations: on the folded lattice the translations
along the chiral vector become pure rotations, while those along
$\vec{a}$ remain pure translations. These elements generate the whole
nanotube from the sector of angle $2\pi/n$ of the elementary cell, with
$2\frac{n_1^2+n_2^2+n_1n_2}{n^2{\cal R}}$ elementary graphene cells. This
number is always greater then 1, pointing out that not all of the
honeycomb lattice translations are taken into account. The missing
translations are neither parallel with nor orthogonal onto $\vec{c}$; on
the rolled up sheet they are manifested as rotations (for fraction of
$2\pi/n$) combined with translations (for fractions of $a$), yielding
the screw axis of the nanotube. Its generator $(C^r_q|\frac{n}{q}a)$
corresponds to the vector
$\vec{z}=r\frac{\vec{c}}{q}+n\frac{\vec{a}}{q}$ of the honeycomb
lattice, which, together with the encountered translations, generates
the whole honeycomb lattice. Thus, $\vec{z}$ can be chosen to form the
elementary honeycomb cell together with the minimal lattice vector
$\vec{c}/n$ along the chiral direction. The honeycomb cell area $S_g$
must be the product of $|\vec{c}|/n$ and the length $na/q$ of the
projection of $\vec{z}$ onto $\vec{a}$: $a|\vec{c}|/q=\sqrt{3}a^2_0/2$.
This gives the order $q$ of the screw axis. Finally, $r$ is found
\cite{YTI} from the condition that projections of $\vec{z}$ on the
$\vec{a}_1$ and $\vec{a}_2$ are coprimes.  This completely determines the
screw axis (${\rm Fr}[x]=x-[x]$ is the fractional part of the rational
number $x$, and $\varphi(m)$ is the Euler function, giving the number of
coprimes less then $m$):
\begin{equation}\label{Escrewstand}
{\bf Z}={\bf T}^r_q,\quad q=2\frac{n^2_1+n_1n_2+n^2_2}{n{\cal R}},\quad
r=\frac{q}{n}{\rm Fr}\left[\frac{n}{q{\cal R}}(3-2\frac{n_1-n_2}{n_1})+
\frac{n}{n_1}\left(\frac{n_1-n_2}{n}\right)^{\varphi(\frac{n_1}{n})-1}\right].
\end{equation}
Especially, for both the zig-zag $(n,0)$ and the armchair $(n,n)$ tubes,
$q=2n$ and $r=1$, i. e. ${\bf Z}={\bf T}^1_{2n}$. Note that $q$ is an even
multiple of $n$. It is equal to the number of the graphene cells in the
elementary cell of the tube $S_t/S_g$.  Therefore, $q/n$ is the number
of the graphene cells in the sector and therefore always greater then 1;
this means that all the single-wall tubes have nonsymorphic symmetry
groups.

To resume, the translational symmetry of the honeycomb lattice appears
as the group ${\bf L}^{(1)}={\bf T}^r_q{\bf C}_n$ of symmetries of the nanotube,
with $q$ and $r$ given by (\ref{Escrewstand}). Its elements
$(C^{rt}_qC^s_n|t\frac{n}{q}a)$ ($t=0,\pm1,\dots$, $s=0,\dots,n-1$)
generate the whole nanotube from any adjacent pair of the nanotube
atoms. The group ${\bf L}^{(1)}$ contains all the symmetries previously
considered in the literature \cite{C-60,WHITE,JISHI,SFCN}. Note that the
screw axis used here is somewhat different to the previously reported
ones, due to the convention \cite{SCREWY}. With this convention $2\pi/q$
is the minimal rotation (followed by some fractional translation) in the
group \cite{YTI}, providing that $q$ is the order of the principle axis
of the isogonal point group. This explains why $q$ is equal to the
number of the graphene cells contained in the elementary cell of the
tube. Note that the translational period $a$ and the diameter $D$ of the
tube are determined by the symmetry parameters $q$ and $n$:
\begin{equation}\label{EaD}
a=\sqrt{\frac{3q}{2{\cal R} n}}a_0,\quad D=\frac{1}{\pi}\sqrt{\frac{{\cal R}
nq}{2}}a_0.
\end{equation}

Besides the translations, there are other symmetries of the honeycomb
lattice: (a) perpendicular rotational axes through the centers of the
hexagons (of order six), through the carbon atoms (of order three) and
through the centers of the edges of the hexagons (of order two); (b) six
vertical mirror planes through the centers of the hexagons formed by the
atoms (or through the atoms); (c) two types of vertical glide planes ---
connecting the midpoints of the adjacent edges, and the midpoints of the
next to nearest neighboring edges of the hexagons.

Among the rotations, only those for $\pi$, leaving invariant the axis of
$\vec{a}$, i. e. the $z$-axis of the tube, remain the symmetry of the
rolled up lattice. Thus, two types of horizontal second order axes
emerge as symmetries of any nanotube (Fig. \ref{Fmed}): $U$, passing
through the center of the deformed nanotube hexagons, and $U'$, passing
through the midpoints of the adjacent atoms. Moreover, the first of
these transformations is obtained when the second one is followed by the
screw axis generator: $U=(C^r_q|\frac{n}{q}a)U'$.  Thus, any of them,
say $U$, complements the principle tube axis ${\bf C}_n$ to the dihedral
point group ${\bf D}_n$. This shows that at least the line group
${\bf T}^r_q{\bf D}_n$ (from the 5th family) is the symmetry group of any
nanotube. Note that $U'$ just permutes the two carbon atoms in the
elementary honeycomb cell, meaning that all the honeycomb atoms are
obtained from an arbitrary one by the translations and the rotation
$U'$. Analogously, the elements of the group ${\bf T}^r_q{\bf D}_n$ generate the
whole nanotube from any of its atoms. The action (\ref{Edefelem}) of the
group elements on the point ${\bf r}_{000}=(\rho_0,\phi_0,z_0)$ (cylindrical
coordinates) gives the points
\begin{equation}\label{Eorbitcoo}
{\bf r}_{tsu}=(C^{rt}_qC^s_nU^u|t\frac{n}{q}a){\bf r}_{000}=
(\rho_0,(-1)^u\varphi_0+2\pi(\frac{t}{q}+\frac{s}{n}),(-1)^uz_0+t\frac{n}{q}a),
\end{equation}
($u=0,1$; $s=0,\dots,n-1$; $t=0,\pm1,\dots$); hereafter, the $x$-axis is
assumed to coincide with the $U$-axis. Using (\ref{EaD}), it can be
shown that the coordinates of the first atom (positioned at
$\frac13(\vec{a}_1+\vec{a}_2)$ on the honeycomb) are
\begin{equation}\label{Efirstcoo}
{\bf r}^C_{000}=(\frac{D}{2},2\pi\frac{n_1+n_2}{nq{\cal R}},
\frac{n_1-n_2}{\sqrt{6nq{\cal R}}}a_0).
\end{equation}
Substituting these values in (\ref{Eorbitcoo}), the coordinates of all
other atoms are obtained.

Rolling up deforms any plane perpendicular onto the graphene sheet,
unless it is either parallel with $\vec{c}$ (then it becomes horizontal
plane) or orthogonal onto $\vec{c}$ (giving vertical plane).  Thus, only
the tubes with the chiral vectors being parallel or orthogonal to the
enumerated mirror and glide planes obtain additional symmetries of these
types. The zig-zag and armchair tubes are immediately singled out by
simple inspection.  Precisely, only in these cases the chiral vector is
in a perpendicular mirror plane; when the sheet is rolled up, this plane
becomes the horizontal mirror plane $\sigma_h$ of the corresponding
nanotubes. Enlarging the previously found point symmetry group ${\bf D}_n$
by $\sigma_h$, the point group ${\bf D}_{nh}$ of the zig-zag and armchair
tubes is obtained. Finally, taking into account the generalized
translations (\ref{Escrewstand}), the full symmetry groups of the
single-wall nanotubes are (besides the factorized notation,
the international symbol is given also):
\begin{mathletters}\label{Elinegroups}
\begin{equation}
{\bf L}_{\mbox{chiral}}={\bf T}^r_q{\bf D}_n=
{\bf L}q_p22,\quad
p=q{\rm Fr}\left[\frac{n{\cal R}}{q}
\frac{(\frac{2n_2+n_1}{n{\cal
R}})^{\varphi(\frac{2n_1+n_2}{n{\cal R}})-1}q-n_2}{2n_1+n_2}\right],
\end{equation}
\begin{equation}
{\bf L}_{\mbox{armchair}}={\bf L}_{\mbox{zig-zag}}={\bf
T}^1_{2n}{\bf D}_{nh}={\bf L}2n_n/mcm.
\end{equation}\end{mathletters}
Their isogonal point groups \cite{IAY} are ${\bf D}_q$ and ${\bf D}_{2nh}$.

{\footnotesize\begin{figure}[hbt]\begin{center}
\unitlength=1.00mm
\linethickness{0.4pt}
\begin{picture}(45.00,20.00)
\put(0.00,0.00){\framebox(45.00,20.00)[cc]{FIG 2}}
 \end{picture}
\caption[]{\label{Ftza} {\bf Symmetries of the single-wall nanotubes:}
{\sl $(8,6)$, $(6,0)$ and $(6,6)$. The horizontal rotational axes $U$
and $U'$ are symmetries of all the tubes, while the mirror planes
($\sigma_v$ and $\sigma_h$), the glide plane $\sigma'_v$ and the
rotoreflectional plane $\sigma'_h$ are symmetries of the zig-zag and
armchair tubes only. The line groups are ${\bf T}^{21}_{148}{\bf D}_2$ for
(8,6), and ${\bf T}^1_{12}{\bf D}_{6h}$ for other two tubes.}}
\end{center}\end{figure}
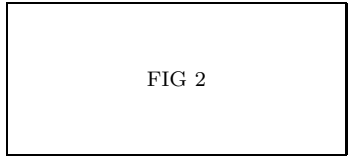}

The line group ${\bf T}^1_{2n}{\bf D}_{nh}$ (13th family) contains various new
symmetries (Fig. \ref{Ftza}), being combinations of the ones mentioned
above. In fact, when $\sigma_h$ has been added to the group
${\bf T}^1_{2n}{\bf D}_n$, the other mirror and glide planes parallel to and
orthogonal onto $\vec{c}$ are automatically included in the symmetry
groups of the zig-zag and armchair nanotubes.  These transformations can
be seen as $\sigma_h$ followed by some of the elements from
${\bf T}^1_{2n}{\bf D}_n$.  At first, there are $n$ vertical mirror planes (one
of them is $\sigma_v=\sigma_hU$, and the others are obtained by pure
rotations; by the previous convention, the $\sigma_h$ plane is the
$xy$-coordinate plane). Bisecting the mirror planes, there are glide
planes (e. g. the product
$(\sigma'_v|\frac12)=(C_{2n}|\frac12a)\sigma_v$), and the vertical
rotoreflection axis of order $2n$ (generated by $\sigma_vU'=
C_{2n}\sigma'_h$, the reflection in the $\sigma'_h$ plane, followed by
the rotation for $\pi/n$).

The vectors obtained from $\vec{c}$ by the rotations from the point
symmetry group ${\bf C}_{6v}$ of the honeycomb lattice, produce the
nanotubes which are essentially the same one, only looked at from the
rotated coordinate systems. Nevertheless, the vertical mirror plane
image of $\vec{c}$ (e. g. in the vertical plane bisecting the angle of
$\vec{a}_1$ and $\vec{a}_2$) produces the tube which can be considered
as the same one only in the coordinate system with the opposite sign
(the coordinate transformation involves the spatial inversion). Thus,
the tubes $(n_1,n_2)$ and $(n_2,n_1)$ are the optical isomers. Only the
mirror image of zig-zag and armchair tube is equivalent to the original,
and these tubes have no optical isomers. Concerning the symmetry groups,
if ${\bf T}^r_q{\bf C}_n$ corresponds to the tube $(n_1,n_2)$, then the
group of the tube $(n_2,n_1)$ is ${\bf T}^{\frac{q}{n}-r}_q{\bf C}_n$
(although isomorphic, these groups are equal only when $q=2n$ and $r=1$,
i. e. only for the zig-zag and the armchair tubes).

\subsection{Double- and multi-wall nanotubes}\label{Ssmwn}
The symmetry of a multi-wall nanotube now can be found as intersection
of the symmetry groups of its single-wall
constituents. This task will be considered for double-wall tubes at
first, and then the results are straightforwardly generalized to the
multi-wall ones. The intersection of the line groups, ${\bf L}={\bf Z}{\bf P}$ and
${\bf L}'={\bf Z}'{\bf P}'$ has the form ${\bf L}_2={\bf Z}_2({\bf
P}\cap{\bf P}')$.  Thus, the intersection of the point groups is looked
for independently of the generalized translations.

As it has been derived in (\ref{Erot}), the tubes $(n_1,n_2)$ and
$(n'_1,n'_2)$ are invariant under the rotations around their axes for
the multiples of the angles $2\pi/n$ and $2\pi/{n'}$ ($n=\text{GCD}(n_1,n_2)$,
$n'=\text{GCD}(n'_1,n'_2)$), respectively. The tube composed of these
coaxially arranged components is invariant under the rotation for
$2\pi/N$, being the minimal common rotation of the components, and its
multiples. Thus, the principle axis subgroup of the double-wall nanotube
is ${\bf C}_N$, with $N=\text{GCD}(n,n')=\text{GCD}(n_1,n_2,n'_1,n'_2)$. The horizontal
second order rotational axis $U$ (and $U'$) is also symmetry of all
single-wall nanotubes. Nevertheless, such an axis remains the symmetry
of the composite tube only if it is common to all of the components, and
then the point symmetry is ${\bf D}_N$.  Obviously, if a nanotube contains
at least one chiral component, then ${\bf D}_N$ is its maximal point
symmetry.  Only the tubes composed exclusively of the zig-zag and
armchair single-wall components may have additional mirror and glide
planes, as well as the rotoreflectional axis. Analogously to the
horizontal axis, these are symmetries of the whole tube only if they are
common for all of the components (the rotoreflectional axis appear only
if the horizontal planes $\sigma'_h$ coincides).

After the point symmetries are thereby completely determined, the more
difficult study of the generalized translational factor ${\bf Z}_2$ remains.
At first, note that it may be completely absent. Suppose that
double-wall tube has the translational period $A$. If the translational
periods of its constituents are $a$ and $a'$, then $A$ is obviously the
minimal distance being multiple both of $a$ and of $a'$: $A=\alpha
a=\alpha'a'$, where $\alpha$ and $\alpha'$ are positive coprimes (to
assure minimality).  Thus, the double-wall tube is translationally
periodic if and only if the translational periods of its constituents
are commensurate, i. e. only when $a'/a$ is rational.  On the contrary,
if $a'/a$ is an irrational number, the composed tube is not
translationally periodic, and ${\bf Z}_2$ is trivial (identical
transformation only); the total symmetry reduces to the already found
point group.

In the commensurate case it remains to examine if the translational
group can be refined by a screw axis, common to all of the single-wall
components. The task is to determine the screw axis generator
$(C^R_Q|F)$ with maximal $Q$, appearing in the both groups
${\bf L}={\bf T}^r_q{\bf C}_n$ and ${\bf L}'={\bf T}^{r'}_{q'}{\bf C}_{n'}$.  Thus, it is
looking for the values of $Q$, $R$ and $F$ (in accordance with
\cite{SCREWY}), such that there exist integers $t$, $s$, $t'$ and $s'$
(enumerating the elements of ${\bf L}$ and ${\bf L}'$) satisfying
\begin{equation}\label{Ebasicscrew}
(C^R_Q|F)=(C^{rt}_qC^s_n|tf)=(C^{r't'}_{q'}C^{s'}_{n'}|t'f')\quad
\mbox{ with }F=\frac{N}{Q}A,\ f=\frac{n}{q}a,\ f'=\frac{n'}{q'}a'.
\end{equation}
Obviously, the fractional translation $F$ is multiple $F=tF^*$ of the
minimal common fractional translation $F^*$, implying
$A=\frac{Q}{N}tF^*$. Analogously to $A$, the translation $F^*$ is found
as the minimal distance being multiple both of $f$ and $f'$; thus it is
given by the unique solution in the coprimes $\phi$ and $\phi'$ of the
equation $F^*=\phi f=\phi'f'$. Since the translational periods of the
single-wall components are multiples of their fractional translations,
$A$ is multiple of $F^*$, i. e. $A=\Phi F^*$.  With help of number
theory, it can be shown \cite{YTI} that only the tubes with the same
${\cal R}$ may be commensurate; then $\alpha
=\phi'=\sqrt{\frac{q'/n'}{\text{GCD}(q/n,q'/n')}}$, $\alpha
'=\phi=\sqrt{\frac{q/n}{\text{GCD}(q/n,q'/n')}}$ and
$\Phi=\sqrt{\frac{qq'}{nn'}}$. Thus, $Q=\Phi N/\tau$, and the minimal
$\tau$ is looked for to provide the finest screw axis. The translational
part of (\ref{Ebasicscrew}) immediately shows that $t=\tau\alpha'$ and
$t'=\tau\alpha$. With this values substituted, the rotational part of
(\ref{Ebasicscrew}) gives the equations:
\begin{equation}\label{Erotscrew}
C^R_Q=C^{r\alpha'\tau}_qC^s_n=C^{r'\alpha\tau}_{q'}C^{s'}_{n'}.
\end{equation}
The minimal $\tau$ for which the last equation is solvable in $s$ and
$s'$ is $\tau=\Phi/\text{GCD}(r\alpha\frac{n'}{N}-r'\alpha'\frac{n}{N},\Phi)$.
Finally,
$Q=N\text{GCD}(r\alpha\frac{n'}{N}-r'\alpha'\frac{n}{N},\sqrt{\frac{qq'}{nn'}})$,
and $R$ is easily found from the first equation (\ref{Erotscrew}).

All these results are immediately generalized to the multi-wall tubes.
Note that the generalized translations and the principle rotational axis
of the multi-wall nanotube depend only on the types of their single-wall
components. On the contrary, the appearance of the mirror and glide
planes and the horizontal axes in the common symmetry group is
additionally determined by the relative positions of these components.

It remains to give the summary of the symmetry groups of the multi-wall
tubes. If at least one of the single-wall constituents is chiral, then
in the commensurate case there are two possibilities: ${\bf T}^R_Q{\bf C}_N$,
corresponding to the general mutual position, and ${\bf T}^R_Q{\bf D}_N$ in the
special mutual positions with common $U$-axis. Analogously, the tube
built of the incommensurate components have the symmetry described by
the point groups ${\bf C}_N$ or ${\bf D}_N$. If the nanotube is built of the
zig-zag and armchair single-wall tubes $(n,0)$ (or $(n,n)$), $(n',0)$
(or $(n',n')$),\dots, the order of the principle rotational axis is
$N=\text{GCD}(n,n',\dots)$. If the tube contains at least one single-wall tube
of both types, no translational periodicity appears and its symmetry is
described by a point group (Tab. \ref{Tsmw}). On the other hand, for the
tube composed of the components of the same type (either zig-zag or
armchair), the translation period is equal to that of the components.
Two different situations may occur: if all the integers $n/N$,
$n'/N$\dots are odd ("odd" case), the translations are refined by the
screw axis ${\bf T}^1_{2N}$; otherwise, if at least one of these integers is
even ("even" case), no screw axis emerges. The analysis of the special
arrangements of the constituents with common horizontal axes, mirror or
glide planes, increasing the symmetry of the total system is summarized
in the table \ref{Tsmw}. Note that according to the various arrangements
of the components, any of the line and axial point groups may be the
resulting symmetry for the commensurate and incommensurate components,
respectively.

Here we give some realistic examples (interlayer distance
\cite{IIJIMA,SAITO+4} of approximately $3.4\mbox{\AA}$, and with diameters
of the single-wall components from $0.7{\rm nm}$ to $3\,{\rm nm}$) of the
double-wall nanotubes, with their symmetry groups in general positions.
With the trivial isogonal point group there are the commensurate tubes
(line group is ${\bf T}{\bf C}_1$) $(6,6)\mbox{--}(11,11)$,
$(7,7)\mbox{--}(12,12)$, $(11,2)\mbox{--}(12,12)$,
$(22,4)\mbox{--}(26,11)$, $(10,0)\mbox{--}(19,0)$,
$(11,0)\mbox{--}(20,0)$, $(7,3)\mbox{--}(14,6)$, $(21,9)\mbox{--}(28,12)$,
$(14,6)\mbox{--}(21,9)$, and the incommensurate pairs (the total symmetry
group is trivial ${\bf C_1}$) $(9,0)\mbox{--}(10,10)$,
$(15,0)\mbox{--}(14,14)$, $(11,2)\mbox{--}(21,0)$,
$(13,4)\mbox{--}(14,14)$, $(10,4)\mbox{--}(19,4)$, $(6,4)\mbox{--}(17,1)$,
$(9,2)\mbox{--}(19,0)$, $(10,0)\mbox{--}(17,3)$, $(5,5)\mbox{--}(17,0)$,
$(6,6)\mbox{--}(19,0)$, $(8,8)\mbox{--}(23,0)$, $(24,9)\mbox{--}(35,6)$,
$(25,7)\mbox{--}(38,0)$, $(27,0)\mbox{--}(31,8)$,
$(17,17)\mbox{--}(30,13)$, $(10,0)\mbox{--}(11,11)$,
$(14,0)\mbox{--}(13,13)$, $(17,0)\mbox{--}(15,15)$,
$(6,4)\mbox{--}(13,7)$, $(8,6)\mbox{--}(21,0)$, $(10,0)\mbox{--}(15,6)$,
$(16,2)\mbox{--}(15,15)$. The commensurate tubes $(9,0)\mbox{--}(17,0)$,
$(9,6)\mbox{--}(15,10)$, $(13,0)\mbox{--}(21,0)$, $(5,5)\mbox{--}(9,9)$,
$(7,7)\mbox{--}(11,11)$, $(11,0)\mbox{--}(19,0)$, have the line group
${\bf T}^1_2{\bf C}_1$, while ${\bf T}{\bf C}_2$ is the symmetry of
$(12,8)\mbox{--}(18,12)$ and CC $(6,4)\mbox{--}(12,8)$. The incommensurate
tubes with the symmetry ${\bf C}_2$ are: $(8,8)\mbox{--}(22,0)$,
$(12,6)\mbox{--}(18,10)$, $(8,14)\mbox{--}(28,0)$,
$(6,6)\mbox{--}(12,10)$, $(16,0)\mbox{--}(14,14)$,
$(22,12)\mbox{--}(28,16)$, $(16,8)\mbox{--}(30,0)$,
$(14,0)\mbox{--}(16,10)$, $(10,8)\mbox{--}(16,12)$,
$(10,2)\mbox{--}(20,0)$, $(26,0)\mbox{--}(30,8)$,
$(22,4)\mbox{--}(22,16)$, $(8,2)\mbox{--}(18,0)$ and
$(8,8)\mbox{--}(16,10)$. Finally, the examples for the higher order
principle axis are commensurate tubes $(5,5)\mbox{--}(10,10)$ (with line
group ${\bf T}^0_4{\bf C}_5$), $(8,8)\mbox{--}(12,12)$ (${\bf T}^0_4{\bf
C}_4$), $(9,0)\mbox{--}(18,0)$, (${\bf T}^0_9{\bf C}_9$),
$(12,0)\mbox{--}(21,0)$, (${\bf T}^0_1{\bf C}_3$), $(14,0)\mbox{--}(22,0)$
(${\bf T}^1_4{\bf C}_2$), and incommensurate ones: $(9,9)\mbox{--}(24,0)$,
$(18,0)\mbox{--}(15,15)$, $(9,3)\mbox{--}(18,3)$, $(12,9)\mbox{--}(27,0)$,
$(15,0)\mbox{--}(18,9)$, $(24,6)\mbox{--}(21,21)$ (all with symmetry ${\bf
C}_3$), $(24,0)\mbox{--}(28,8)$, $(20,12)\mbox{--}(32,8)$,
$(28,0)\mbox{--}(32,8)$ (${\bf C}_4$), $(15,15)\mbox{--}(35,0)$ (${\bf
C}_5$), $(7,7)\mbox{--}(21,0)$ (${\bf C}_7$), $(11,0)\mbox{--}(11,11)$
(${\bf C}_11$), $(12,0)\mbox{--}(12,12)$ (${\bf C}_12$) and
$(13,0)\mbox{--}(13,13)$ (${\bf C}_13$).

\normalsize\begin{table}[hbt]\begin{center}{\footnotesize
\caption[]{\label{Tsmw}{\footnotesize
{\bf Symmetry of the multi-wall zig-zag and armchair tubes.} For the
periodic tubes, the line groups (and families) and the
isogonal groups are in the "odd" columns if all the ratios $n/N$,
$n'/N$, \dots are odd, and in the "even" columns
otherwise. The point groups of the tubes with both zig-zag and armchair
components is in the last
column. In the first column the relative positions of the component tubes
are characterized by the coinciding symmetry elements (beside the
common principle axis in the general position).  Here, $(U,U')$ denotes
the horizontal axis, which is $U$-axis in some of the constituents, and
$U'$-axis in the remaining ones (to exclude the additional mirror or
glide planes). Also, $(\sigma_h,\sigma'_h)$ is the plane being
$\sigma_h$ in some of constituents (with even $n$, necessarily), and
$\sigma'_h$ in the remaining tubes; in the incommensurate case, the same
groups are obtained when $\sigma'_h$ planes are in common.\\}}
\begin{tabular}{c|cccccc|c}
Relative
&\multicolumn{4}{c}{Line group}&\multicolumn{2}{c|}{Isogonal group}&
   Point\\ 
position
&\multicolumn{2}{c}{"Odd"}&\multicolumn{2}{c}{"Even"}&"Odd"&"Even"&
  group\\ \hline
General
  &${\bf T}^1_{2N}{\bf C}_N$&(1)&${\bf T}{\bf C}_N$ &(1)&${\bf C}_{2N}$&${\bf C}_N$&
  ${\bf C}_N$\\ 
$\sigma_h$
  &${\bf T}^1_{2N}{\bf C}_{Nh}$&(4)&${\bf T}{\bf C}_{Nh}$ &(3)&${\bf C}_{2Nh}$&${\bf C}_{Nh}$&
  ${\bf C}_{Nh}$\\ 
$\sigma_v$
  &${\bf T}^1_{2N}{\bf C}_{Nv}$&(8)&${\bf T}{\bf C}_{Nv}$ &(6)&${\bf C}_{2Nv}$&${\bf C}_{Nv}$&
  ${\bf C}_{Nv}$\\ 
$\sigma'_v$
  &${\bf T}^1_{2N}{\bf C}_{Nv}$&(8)&${\bf T}_c{\bf C}_{Nv}$&(7)&${\bf C}_{2Nv}$&${\bf C}_{Nv}$&
  ${\bf C}_{N}$\\ 
$(U,U')$
  &${\bf T}^1_{2N}{\bf D}_{N}$&(5)&${\bf T}{\bf D}_{N}$ &(5)&${\bf D}_{2N}$&${\bf D}_{N}$&
  ${\bf D}_{N}$\\ 
$\sigma_h,\sigma_v$
  &${\bf T}^1_{2N}{\bf D}_{Nh}$&(13)&${\bf T}{\bf D}_{Nh}$&(11)&${\bf D}_{2Nh}$&${\bf D}_{Nh}$&
  ${\bf D}_{Nh}$\\ 
$\sigma_h,\sigma'_v$
  &${\bf T}^1_{2N}{\bf D}_{Nh}$&(13)&${\bf T}_c{\bf C}_{Nh}$&(12)&${\bf D}_{2Nh}$&${\bf D}_{Nh}$&
  ${\bf C}_{Nh}$\\ 
$(\sigma_h,\sigma'_h)$
  &${\bf T}^1_{2N}{\bf C}_{Nh}$&(4)&${\bf T}{\bf S}_{2N}$&(2)&${\bf C}_{2Nh}$&${\bf S}_{2N}$&
  ${\bf S}_{2N}$\\ 
$(\sigma_h,\sigma'_h)$, $\sigma_v$
  &${\bf T}^1_{2N}{\bf D}_{Nh}$&(13)&${\bf T}{\bf D}_{Nd}$&(9)&${\bf D}_{2Nh}$&${\bf D}_{Nd}$&
  ${\bf D}_{Nd}$\\ 
$(\sigma_h,\sigma'_h)$, $\sigma'_v$
  &${\bf T}^1_{2N}{\bf D}_{Nh}$&(13)&${\bf T}_c{\bf S}_{2N}$&(10)&${\bf D}_{2Nh}$&${\bf D}_{Nd}$&
  ${\bf S}_{2N}$\\ 
\end{tabular}}\end{center}\end{table}\normalsize

\section{Concluding remarks}\label{Sconc}
All the geometrical symmetries of the nanotubes are found. In addition
to the rotations, translations and screw-axes, observed previously, the
single-wall tubes always possess horizontal rotational axes; the zig-zag
and armchair tubes have mirror and glide planes in addition. Thus, the
full symmetry group is ${\bf T}^r_q{\bf D}_n$ for single-wall chiral tubes
and ${\bf T}^1_2{\bf D}_{nh}$ for zig-zag and armchair ones. The
parameters $q$ and $r$ of the helical group are found in the simple and
closed form. Since $2\pi/q$ is the angle of the minimal rotation (combined
with fractional translation) performed by the symmetry group, the order of
the principle axis of the isogonal group is $q$ and it is always even.
Moreover, $2q$ is the number of the carbon atoms in the elementary
translational cell of the tube. Let us only mention here that the
different tubes cannot have the same symmetry parameters $q$, $r$,
$n$ and $a$. This profound property means that the line group is
sufficient to reconstruct the tube (as it is demonstrated by
(\ref{Eorbitcoo})), i. e. that the symmetry completely determines the
geometry and all consequent characteristics of nanotube.  The symmetries
of the multi-wall tubes are quite diverse: depending on types of the
single-wall components and their arrangements, all the line and
axial-point groups emerge: armchair and zig-zag tubes can be combined to
make a prototype for any line or axial symmetry group. This immediately
shows that the properties of the nanotubes may vary greatly, depending not
only on the single-wall constituents, but also on their mutual positions.

There are many physical properties based on symmetry, and the presented
classification of the nanotubes according to their symmetry can be
widely exploited. The most familiar consequence of symmetry, the special
forms of the tensors related to the characteristics of the system,
depends on the isogonal group.

Further, the symmetry can be used to find good quantum numbers.  To
begin with the single-wall nanotubes. The translational periodicity is
reflected in the conserved quasi-momentum $k$, taking the values from
the 1D Brillouine zone $(-\pi,\pi]$, or its irreducible domain
\cite{ALTMAN2} $[0,\pi]$.  Also, the $z$-component of the quasi-angular
momentum $m$ is the quantum number caused by the symmetry of the
principle rotational axis; it takes on the integer values from the
interval $(-\frac{n}{2},\frac{n}{2}]$, and characterizes the nanotube
quantum states.  The parity with respect to reversal of the $z$-axis,
induced by the horizontal rotational axis $U$, is the last quantum
number common to all the single-wall tubes.  The even and the odd states
with respect to this parity are conventionally denoted by $+$ and $-$.
For the zig-zag and the armchair tubes there is additional vertical
mirror plane parity, introducing the quantum numbers $A$ and $B$, to
distinguish between the even and the odd states (the parity with respect
to the horizontal mirror plane is dependent on the above discussed $U$
and $\sigma_v$ parities, $\pm$ and $A/B$). Concerning the multi-wall
tubes, $m$ is quantum number again. Again, $z$-reversal and vertical
mirror parities may appear, depending on the concrete symmetry of the
nanotube. Nevertheless, the tubes with incommensurate components are not
periodic, and in such cases the quasi-momentum $k$ is not an appropriate
quantum number; it may be interesting experimental question whether the
approach of modulated systems can be applied to restore this quantity.
The simple criterion of commensurability of the single-wall tubes is
derived: they have same ${\cal R}$ and $\sqrt{\frac{qq'}{nn'}}$ is an
integer. The involved symmetry parameters $q$ and $n$ are discrete,
allowing exact experimental check of commensurability.

The enumerated quantum numbers may be used to discuss and predict many
characteristics of the nanotubes, but the most sophisticated approach to
classification and properties of different quantum states is based on
the irreducible representations of the corresponding line
\cite{IBB,IY-93} and point groups. Let us remind that these
representations are labeled by the derived quantum numbers. The most
exhaustive possible information on selection rules, comprising the
conservation of quantum numbers, for the processes in the nanotubes has
become available \cite{YIBB} after the full line (or point) group
symmetry has been established.

The dimension of an irreducible representation equals degeneracy of the
corresponding energy level. For the periodic tubes, the degeneracy of
the energy bands is at most fourfold; nevertheless, if the time
reversal symmetry of the (spin-independent) Hamiltonian is encountered,
the maximal degeneracy is eight-fold \cite{YI-CO}. Further, the possible
degeneracies are only two-, four- and eight-fold. As for the multi-wall
nanotubes with incommensurate components, the dimensions of the
irreducible representations of the axial point groups are one, two and
(if the time reversal symmetry is included) four, showing the possible
degeneracies of the energy levels. Note that the maximal of the
enumerated degeneracies (eight- and four-fold) is not possible for the
tubes containing at least one chiral single-wall component. Moreover,
the degeneracy of the multi-wall tube in the general position of its
component is at most two-fold, which is caused by the time reversal
symmetry exclusively.

Also, the lattice dynamics can be studied. As it has been mentioned
above, the whole single-wall tube can be obtained from its arbitrary
atom by the action of the elements of its symmetry group
(\ref{Elinegroups}); in group theoretical
language, this means that the whole nanotube is a single orbit of this
group \cite{IRY}, and this is the orbit $a_1$ for the chiral, $b_1$ for
the zig-zag, and $d_1$ for the armchair tubes \cite{STABY}. Thus, the
normal modes (phonons), are already classified \cite{IY-93}. The dynamical
representation of the chiral tube is decomposed onto the following
irreducible components (the summation over $k$ is over the interval
$(0,\pi)$; in the primed sum $m$ takes integer values from $(0,{n\over
2})$, otherwise from $(-{n\over 2},{n\over 2}]$; the components with
$m=n/2$ appear only for $n$ even):
$$D^{\rm dyn}_{\rm chiral}=3({{}_oA^+_o} +{{}_oA^-_o}
+{{}_{\pi}A^+_o}+{{}_{\pi}A^-_o}+{{}_oA^+_{n/2}}+{{}_oA^-_{n/2}}+
{{}_{\pi}A^+_{n/2}}+{{}_{\pi}A^-_{n/2}})+ 6{\sum_m}'({{}_{\pi}E_m}
+{{}_oE_m})+6{\sum_{k,m}}{{}_kE_m}.$$
It can be seen that all of the $6n$ vibrational bands are double
degenerate, as it has been anticipated.  As for the zig-zag and armchair
tubes the corresponding decompositions are (summation in $m$ is over
integers from $(0,n)$, and in the primed sums from $(0,{n\over 2})$):
$$D^{\rm dyn}_{{\rm zig-zag}}=2({{}_oA^+_o}+{{}_oA^-_o}+{{}_oA^+_n}
+{{}_oA^-_n}+{{}_{\pi}E_B})+ {{}_oB^+_o} + {{}_oB^-_o}
+{{}_oB^+_n}+{{}_oB^-_n}+4{{}_{\pi}E_A}+$$
$$3{\sum_m}({{}_oE^+_{m}}+{{}_oE^-_{m}})+2{\sum_k}[
{{}_kE_{B_o}}+{{}_kE_{B_n}}+2({{}_kE_{A_o}}
+{{}_kE_{A_n}})]+6{\sum_m}'{{}_{\pi}G_{m}}+6{\sum_{k,m}}{{}_kG_{m}}
+3({{}_{\pi}E^+_{n/2}}+{{}_{\pi}E^-_{n/2}}),$$
$$D^{\rm dyn}_{\rm armchair}=2({{}_oA^+_o} +{{}_oB^+_o} +{{}_oA^+_n}
+{{}_oB^+_n})+{{}_oA^-_o}+{{}_oB^-_o}+{{}_oA^-_n}+{{}_oB^-_n}+3({{}_{\pi}E_A}+
{{}_{\pi}E_B})+6{\sum_{k,m}}{{}_kG_{m}}$$
$$+3{\sum_k}({{}_kE_{A_o}}+{{}_kE_{B_o}}+{{}_kE_{A_n}}+
{{}_kE_{B_n}})+{\sum_m}( 4{{}_oE^+_{m}}+2{{}_oE^-_{m}}) +
6{\sum_m}'{{}_{\pi}G_{m}}+3({{}_{\pi}E^+_{n/2}}+{{}_{\pi}E^-_{n/2}}).$$
Analogous data can be directly found for each nanotube. This
classification can be used to simplify calculation of the vibrational
bands \cite{CHARLIER}; the obtained bands are automatically labelled by
the symmetry based quantum numbers, meaning that Raman and IC spectra can
be directly extracted by the selection rules.  Note in this context, that
the Jahn-Teller theorem is proved both for the point and for the line
groups \cite{IY-93}.

Bo\v zovi\'c (OXXEL GmbH, Bremen) and dr G. Bicz\'o (Central Research
Institute for Chemistry, Budapest), who paid our attention to this
subject. Also, we thank to dr R. Kosti\'c (Institute of Physics, Beograd)
for some remarks.



\begin{references}
\bibitem{SCREWY}
The standard notation is used: $\sigma_v$ is vertical mirror plane, $I$
is $3\times 3$ identity matrix, while $a$ denotes the translational
period.  Also, $C^r_q$ is the rotation for $2\pi r/q$ around $z$-axis,
where $0\leq r<q/n$ and $q/n$ are coprime integers.  Instead of this
helical group ${\bf Z}$, generated by  $(C^r_q|\frac{n}{q}a)$, the groups
generated by $(C^{r_i}_q|\frac{n}{q}a)$, with $r_i=r+i\frac{q}{n}$,
$i=1,\dots,n-1$, give the same full symmetry group \cite{WHITE}.
Among these $r_i$ there is at least one being coprime with $q$; with this
$r_i$, it is obvious that the set $(C^{tr_i}_q|t\frac{n}{q}a)$ contains
the rotations for all the multiples of $2\pi/q$ (followed by some
translations).
\end{references}
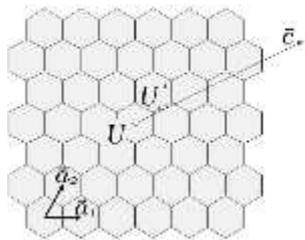
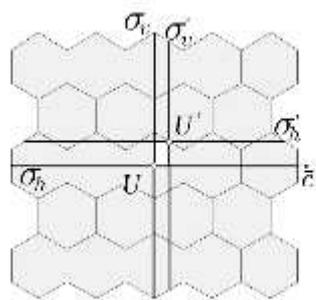
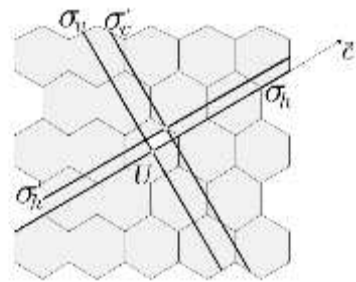
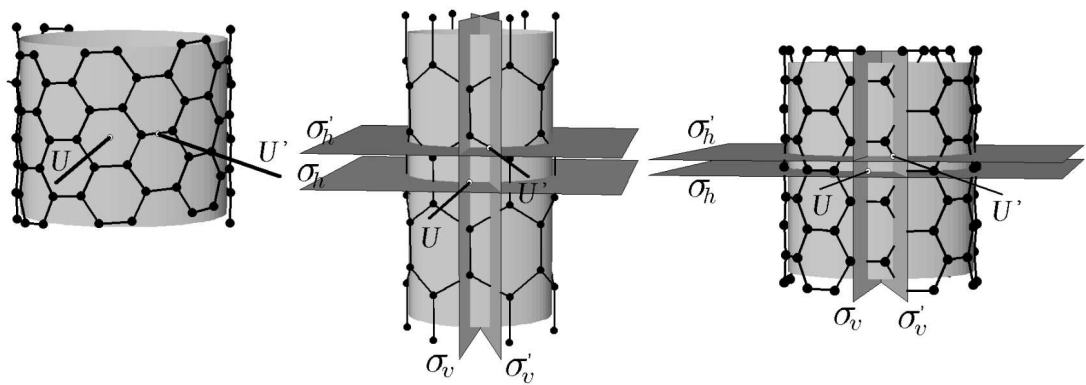
\end{document}